\documentclass[aps,twocolumn,prb]{revtex4}
\usepackage{amsmath,revsymb}
\usepackage[xdvi]{graphicx}
\usepackage{dcolumn}

\begin{document}

\title{Electronic structure of BAs and boride III--V alloys}
\author{Gus L. W. Hart}
\author{Alex Zunger}
\affiliation{National Renewable Energy Laboratory, Golden, CO 80401}
\date{\today}
\begin{abstract}
Boron arsenide, the typically-ignored member of the III--V arsenide
series BAs--AlAs--GaAs--InAs is found to resemble silicon
electronically: its $\Gamma$ conduction band minimum is $p$-like
($\Gamma_{15}$), not $s$-like ($\Gamma_{1\text{c}}$), it has an
$X_{1\text{c}}$-like indirect band gap, and its bond charge is
distributed almost equally on the two atoms in the unit cell,
exhibiting nearly perfect covalency. The reasons for these are tracked
down to the anomalously low atomic $p$ orbital energy in the boron
and to the unusually strong $s$--$s$ repulsion in BAs relative to most
other III--V compounds. We find unexpected valence band offsets of BAs
with respect to GaAs and AlAs. The valence band maximum (VBM) of BAs
is significantly higher than that of AlAs, {\em despite} the much
smaller bond length of BAs, and the VBM of GaAs is only slightly
higher than in BAs. These effects result from the unusually strong
mixing of the cation and anion states at the VBM. For the BAs--GaAs
alloys, we find (i) a relatively small ($\sim$3.5 eV) and
composition-independent band gap bowing. This means that while
addition of small amounts of nitrogen to GaAs {\em lowers} the gap,
addition of small amounts of boron to GaAs {\em raises} the gap (ii)
boron ``semi-localized'' states in the conduction band (similar to
those in GaN--GaAs alloys), and (iii) bulk mixing enthalpies which are
smaller than in GaN--GaAs alloys. The unique features of boride III--V
alloys offer new opportunities in band gap engineering.

\end{abstract}


\maketitle

\section{Introduction}

With the advent of state-of-the-art techniques for growing
semiconductor alloys on common substrates such as GaAs, silicon, and
germanium, semiconductor compounds which previously were very
difficult to synthesize are now routinely achieved. Techniques such as
metal-organic chemical vapor deposition (MOCVD), molecular beam
epitaxy (MBE), and pulsed laser ablation (PLD) have provided the
opportunity to synthesize and study a large number of nitride,
phosphide, and antimonide semiconductor alloys. Of particular recent
interest are alloys of a {\em wide gap} semiconductor (e.g. nitrides)
with a ``conventional'' III--V semiconductor because of their promise
in optical applications. Two diverging scenarios were considered: (i)
using a significant amount (10\%--30\%) of the wide gap component to
shift the alloy band gap to the {\em blue} (e.g., adding $\sim$20\%
GaN to InN) for light-emitting diode or laser applications, and (ii)
using a {\em small} amount of the wide gap semiconductor to shift the
alloy band gap to the {\em red} (e.g., adding 1--3\% GaN to GaAs) for
photovoltaic applications. The latter effect occurs naturally if the
band gap bowing parameter $b$ is larger than the difference of the
band gaps of the constituents (e.g., ZnS-ZnTe; GaAs-GaN). In this case
addition of small amounts of the wide gap components acts to initially
lower the band gap of the small gap component. For example, one can
achieve the technologically desired 1 eV gap if one adds nitrogen to
GaAs or to InGaAs.

When boron is substituted into GaAs, it can go to either a gallium
site or an arsenic site. Normally boron prefers isovalent substitution
on the gallium site,\cite{addinall,newman,maguire} which is the case
we study here. In the other case when boron goes to the arsenic site
(a boron ``antisite'' defect), the boron acts as an acceptor and this
antisite defect has been the subject of numerous
studies.\cite{woodhead,gledhill,chadi,tischler,hoke} Growth conditions
determine whether boron goes to the gallium site as an isovalent
substition or to the arsenic site as an acceptor. For example,
B$_{As}$ antisite defects are known to occur in Ga-rich samples of
GaAs grown by the liquid encapsulated Czochralski (LEC) method, but
for GaAs crystals taken from stoichiometric or As-rich melts,
electrically active boron or boron complexes are not found,\cite{ta}
indicating that the boron atoms have substituted isovalently to the
gallium sites. Not as much is known about the {\em epitaxial} growth
conditions under which isovalent or antisite boron incorporation
occurs. It is reasonable to suppose that B$_{As}$ antisite defects
will be more likely under Ga-rich conditions and that As-rich growth
conditions will lead to isovalent boron incorporation, similar to the
case for LEC-grown GaAs. In this study, we focus on isovalent BGaAs
alloys where boron occupies gallium sites.

In this paper we will explore BAs as an alternative to GaN as a wide
gap partner for alloying conventional III--Vs such as AlAs and
GaAs. Very little is known about BAs: As early as 1966, Ku mentioned
the possibilities of BAs--GaAs as a boride III--V alloy with
potentially useful properties.\cite{ku,sakai} However, difficulties in
fabricating BAs or simple solid solutions of zinc-blende
BAs--GaAs\cite{ku,vorobev,chu,perri,williams} prior to the development
of current epitaxial techniques have prevented a careful study of the
alloy properties. Now that epitaxial techiques have eased these
difficulties somewhat, boride semiconductor alloys are generating
renewed interest.\cite{tischler,hoke,manasevit,timmons,geisz} There
are a number of theoretical studies that examine different aspects of
pure
BAs\cite{stukel,wentzcovitch,wentzcovitch2,prasad,ferhat,surh,benkabou,bouhafs}
but theoretical studies of the boron {\em alloys} are lacking.

Important questions that one would want to answer include:

(i) How will alloying with BAs affect the band gaps and other
properties of GaAs or AlAs?

(ii) To what extent does BAs fit the well-known trends in band gaps,
band offsets and bonding patterns of the arsenide III--V series
BAs--AlAs--GaAs--InAs?

(iii) Will the band gap bowing of B$_x$Ga$_{1-x}$As alloys be as
anomalous as the (very large and composition-dependent) bowing in
GaAs$_{1-x}$N$_x$?

(iv) Will cation substitution by boron lead to unusual wavefunction
localization effects found to exist for anion substitition by
nitrogen?\cite{mattila,bellaiche,shw-bowing}

Two features of boron make boride semiconductors fundamentally
different from common III--V or II--VI semiconductors. The first is
that, like nitrogen, boron is in the first row of the periodic table
and has deep $p$ orbitals and a small atomic size. The second feature
is, unlike nitrogen, boron has a low electronegativity. This leads to
highly covalent compounds, unlike nitride semiconductors, which have a
strong ionic character. This paper examines (i) zinc-blend BAs and its
place in the III--As family of semiconductors and (ii) boron
substitution of gallium in GaAs, including alloy bowing, band offsets
and mixing enthalpies. Our main findings are:

{\em Zinc-blende BAs:} Surprisingly, we find that, electronically, BAs
resembles silicon rather than other III--V semiconductors. Similar to
silicon and in contrast to most III--Vs, the lowest Brillouin-zone
center conduction band of BAs has $p$ symmetry ($\Gamma_{15\text{c}}$)
rather than $s$ symmetry ($\Gamma_{1\text{c}}$), and, like silicon,
its total valence charge density shows almost symmetric distribution
of charge around the two atoms in the unit cell. The reasons for the
silicon-like conduction band ordering in BAs are: (i) the small
repulsion of the bonding and antibonding $p$ states due to the low $p$
orbital energy of boron, as well as the unusual hybridization of both
cation and anion $p$ states at the VBM, and (ii) the repulsion of the
cation and anion $s$ states which is much stronger in BAs than in
AlAs, GaAs, and InAs. As a result of the $p$--$p$ hybridization
(covalent bonding) mentioned in (i), we also find that the valence
band offset of BAs relative to other members of the III--As family is
unusually high.

{\em BAs--GaAs alloys:} (i) The band gap bowing is relatively small
($\sim$3.5 eV) and composition-independent, in stark contrast to
GaN--GaAs alloys. Because of this small bowing, the addition of BAs to
GaAs {\em increases} the gap, thus, unlike nitrogen, addition of boron
into GaAs or InGaAs will not lead to the desired 1 eV material. (ii)
The lower energy conduction band states are ``semi-localized'' states
around the boron atoms, e.g., the CBM is strongly localized near the
boron but extended at longer distances while the VBM is completely
delocalized. (iii) The bulk mixing enthalpy of BAs in GaAs is much
lower than that of GaN in GaAs, indicating that the bulk solubility of
boron in III--V compounds may be higher than that of nitrogen and thus
higher composition ranges may be possible with the boride
alloys. These findings indicate that boride III--V alloys provide
new opportunities in band gap engineering.
\section{Methods of Calculation}
\label{details}
\subsection{The LAPW calculations}

We used density functional theory within the local density
approximation (LDA),\cite{kohn} as implemented by the full-potential
linearized-augmented-planewaves (LAPW) method\cite{singh1,singh2}
(\textsc{wien97} implementation\cite{blaha}). The exchange-correlation
potential of Perdew and Wang was used.\cite{perdew} In the
calculations with less than 32 atoms, the plane wave kinetic energy
cutoff for the expansion in the interstitial region was 16 Ryd
(approximately 130 basis function per atom). The muffin tin (MT) radii
were 1.65 bohr for boron and 2.2 bohr for arsenic, aluminum, gallium,
and indium. In the large supercell calculations (32 or more atoms), a
slightly smaller plane wave kinetic energy cutoff 13 Ryd
(approximately 120 basis functions per atom) and a larger boron MT
radius of $1.8$ bohr (2.1 for arsenic, aluminum, gallium, and indium)
was used to ease the computational burden of the larger cells. Our
convergence studies indicate that the error in the individual
eigenstates is less than 5 meV for the valence and lower conduction
bands. The calculations were run until the variation in the total
energy between several self-consistency cycles was $< 10^{-5}$
Ryd. The experimental lattice constants were used in all the
calculations of the individual compounds. The experimental lattice
constants are 4.777 \AA, 5.660 \AA, 5.653, and 6.058 \AA\ for BAs,
AlAs, GaAs, and InAs, respectively.

The $k$ point mesh used in the calculation of the simple binary
compounds (zinc-blende BAs, AlAs, GaAs, and InAs) was a $4\times
4\times 4$ mesh of Monkhorst and Pack special points (10 points in the
irreducible wedge of the Brillouin zone [BZ]).\cite{monkhorst} The
superlattice calculations for the valence band offsets as well as the
supercell calculations for the alloy studies used $k$ point meshes
equivalent\cite{froyen} to the $4\times 4\times 4$ mesh used in the
calculation of the simple zinc-blende binary compounds. Using
equivalent $k$ point meshes is particularly important for calculations
such as enthalpies of formation in order to eliminate uncertainties
due to the statistical errors of different $k$ point meshes. Thus, one
can use much smaller $k$ point meshes to achieve the required
accuracies than would otherwise be necessary.

\subsection{Partial DOS, band characters, and valence charge density}
It is useful to analyze the orbital character of different states. The
band character (or orbital population)
$Q_{\ell}^{(\alpha)}(\epsilon,\mathbf{k})$ is the $\ell\text{-th}$
angular momentum component of the charge due to wavefunction
$\psi(\epsilon,\mathbf{k})$ enclosed in a sphere $\Omega_{MT}$ of
radius $R_{MT}^{(\alpha)}$ about atom $\alpha$
\begin{equation}
\label{ql}
Q_{\ell}^{(\alpha)}(\epsilon,{\mathbf k})=
\int 
_{\Omega_{MT}}
|\hat{P}_{\ell}\,
\psi(\epsilon,{\mathbf k}, {\mathbf r})|^2
\,d{\mathbf r}
\end{equation}
where $\hat{P}_{\ell}$ is an angular momemtum projection operator with
origin at site $\alpha$. Because the interstitial region outside the
muffin tin spheres is excluded when the band characters are
calculated, the muffin tin radii were chosen to match the rationalized
tetrahedral radii of Phillips.\cite{phillips} 

The partial DOS are determined by integrating
$Q^{(\alpha)}_{\ell}(\epsilon,\mathbf{k})$ over all $\mathbf{k}$ in
the Brillouin zone (BZ);
\begin{equation}
\label{eq:dos}
N^{(\alpha)}_{\ell}(\epsilon)=\int_{BZ}Q^{(\alpha)}_{\ell}
(\epsilon,{\mathbf{k}})\,d\mathbf{k}.
\end{equation}
The angular-momentum decomposition of the total valence charge can
then be determined by summing $N_{\ell}(\epsilon)$ over the valence
bands;
\begin{equation}
\label{eq:qtot}
q^{\text{tot}}_{\ell}=
\int^{\text{VB$_{\text{max}}$}}_{\text{VB$_{\text{min}}$}}
N_{\ell}(\epsilon) \,d\epsilon.
\end{equation}

{\em Valence charge densities:} The charge density is constructed from
the highest $N_B$ occupied bands as
\begin{equation}
\label{eq:VBrho}
\rho_{\text{val}}({\mathbf r})=
\sum_{n=1}^{N_B}
\int_{BZ}
|\psi_{n,{\mathbf k}}({\mathbf r})|^2 \,d{\mathbf{k}}.
\end{equation}
For plotting $\rho_{\text{val}}$ we use $N_B=4$. Thus, cation $d$ bands
are not included in the construction.

The ``valence deformation density'' $\Delta \rho_{\text{val}}({\mathbf
r})$ describes the difference between the solid-state density and a
superposition of densities of spherical atoms in their ground
state, 
\begin{equation}
\label{eq:rhodiff}
\Delta \rho_{\text{val}}({\mathbf r})=
\rho_{\text{val}}({\mathbf r})-
\rho_{\text{sup}}({\mathbf r}).
\end{equation}
The valence charge difference is defined as the difference
between the crystal valence density (including the cation $d$ bands,
when occupied) and a superposition of free, spherical LDA atomic
densities arranged in the configuration of the crystal. The atomic
configuration of the free atoms are $d^{10}s^2p^1$ for the cations and
$s^2p^3$ for the anions.

\subsection{Band offsets}
The offset $\Delta E_v(AX/BY)$ between the valence
band maxima of two semiconductor compounds $AX$ and $BY$ forming a
heterostructure is calculated using a method similar to that
used in photoemission spectroscopy.\cite{vbo} 
\begin{equation}
\label{vboeqn}
E_v(AX/BY)=\Delta
E^{BY}_{v,C'}-\Delta E^{AX}_{v,C}+\Delta E^{AX/BY}_{C,C'},
\end{equation}
where
\begin{eqnarray}
\Delta E^{AX}_{v,C}&=&E^{AX}_v-E^{AX}_C\nonumber\\
\Delta E^{BY}_{v,C'}&=&E^{BY}_v-E^{AX}_{C'}
\end{eqnarray}
are the energy seperations between the core levels ($C$ and $C'$) and
the valence band maximums for the pure $AX$ and $BY$ compounds. The
third term in equation (\ref{vboeqn}),
\begin{equation}
\Delta E^{AX/BY}_{C,C'}=E^{BY}_{C'}-E^{AX}_{C}
\end{equation}
is the difference in the core level binding energy between the two
compounds $AX$ and $BY$ in the $AX/BY$ heterojunction. To calculate
the ``natural band offset'', $\Delta E^{AX}_{v,C}$ and $\Delta
E^{BY}_{v,C'}$ are calculated for $AX$ and $BY$ at their cubic,
equilibrium lattice constants. The core level difference $\Delta
E^{AX/BY}_{C,C'}$ is obtained from calculations of $(AX)_n/(BY)_n$
superlattices in the (001) direction. The period $n$ needs to be large
enough so that $AX$-like and $BY$-like properties can be identified
far from the interface. For the core states (denoted $C$ and $C'$), we
use the anion $1s$ states. Table \ref{offsets} shows the dependence of
the valence band offset calculations on the period of the superlattice
and on which core states are used. Because of the large size mismatch
of BAs and GaAs, at least $n=4$ is required to sufficiently converge
$\Delta E^{AX/BY}_{C,C'}$. We also find that while a similar offset is
obtained for AlAs/GaAs when either anion or cation core levels are
used, for the borides only anion core levels provide a rapidly
convergent band offset with respect to the superlattice period $n$.

\begin{table}[tbp] \centering
\caption{Convergence test for the valence band offsets of BAs/GaAs,
BAs/AlAs, and AlAs/GaAs using two different superlattice periods.
Units are eV. Band offsets were calculated using both anion and cation
core states [see Eq.~(\ref{vboeqn})]. Note that while for
AlAs/GaAs a similar offset is obtained when either anion or cation
core levels are used, for the borides only anion core levels provide a
rapidly convergent band offset with respect to the superlattice period
$n$.}
\label{offsets} 
\begin{ruledtabular}
\begin{tabular*}{\linewidth}{@{\extracolsep{\fill}}lddd}
Superlattice &
\multicolumn{1}{c}{Using} & \multicolumn{1}{c}{Using}
 & \multicolumn{1}{c}{Using} \\ 
&\multicolumn{1}{c}{1s anion} & \multicolumn{1}{c}{2s anion}
 & \multicolumn{1}{c}{1s cation} \\ 
\hline
(BAs)$_2$/(GaAs)$_2$  & 0.18 &  0.11 & 0.83 \\ 
(BAs)$_4$/(GaAs)$_4$  & 0.19 &  0.12 & 0.58 \\ 
\hline
(BAs)$_2$/(AlAs)$_2$  &-0.38 & -0.45 & 0.58 \\
(BAs)$_4$/(AlAs)$_4$  &-0.39 & -0.47 & 0.17 \\
\hline
(AlAs)$_2$/(GaAs)$_2$ & 0.50 &  0.50 & 0.41 \\ 
(AlAs)$_4$/(GaAs)$_4$ & 0.51 &  0.51 & 0.47 \\ 
\end{tabular*}
\end{ruledtabular}
\end{table}

\subsection{Alloy enthalpies of mixing, alloy bowing, and equivalent
$k$ points}
\label{d:deltaH}

\begin{table*}[t] \centering
\caption{Definition of the supercells used in this study and
equivalent $k$ points. Lattice vectors are given in units of $a_0$ and
equivalent $k$ points are given as fractions of the reciprocal lattice
vectors.}
\label{kpoints} 
\squeezetable
\begin{ruledtabular}
\begin{tabular*}{\linewidth}{@{\extracolsep{\fill}}lccc}
System & Lattice vectors & Equivalent $k$ points & Relative weight \\ 
\hline
$AC$, $BC$ zinc-blende  & (1/2, 1/2,   0) & (0, 0, 1/8)    &1\\
                      & (1/2,   0, 1/2) & (0, 0, 3/8)    &1\\
                      & (0, 1/2, 1/2)   & (0, 1/8, 3/4)  &3\\
                      &                 & (0, 1/8, 1/4)  &3\\
                      &                 & (0, 1/8, 1/2)  &3\\
                      &                 & (0, 1/4, 5/8)  &3\\
                      &                 & (1/8, 1/4, 1/2)&6 \\
                      &                 & (0, 1/4, 3/8)  &3\\
                      &                 & (0, 3/8, 1/2)&3\\
                      &                 & (1/8, 3/8, 5/8)  &6\\
\hline
$(AC)_1$/$(BC)_1$ (100) superlattice  & (1/2, 1/2, 0) & (0, 1/8, 1/8) & 1\\
                          & (1/2, -1/2, 0)& (0, 1/8, 3/8) &1\\
                          & (0, 0, 1)     & (0, 3/8, 1/8) &1\\ 
                      &                 & (0, 3/8, 3/8) &1\\
                      &                 & (1/8, 1/4, 1/8) &2\\
                      &                 & (1/8, 1/4, 3/8) &2\\
                      &                 & (1/8, 1/2, 1/8) &1\\
                      &                 & (1/8, 1/2, 3/8) &1\\
                      &                 & (1/4, 3/8, 1/8) &2\\
                      &                 & (1/4, 3/8, 3/8) &2\\
                      &                 & (3/8, 1/2, 1/8) &1\\
                      &                 & (3/8, 1/2, 3/8) &1\\
\hline
$(AC)_2$/$(BC)_2$ (100) superlattice  & (1/2, 1/2, 0) &(0, 1/8, 1/4)  &1\\
                          & (1/2, -1/2, 0)     & (0, 3/8, 1/4)  &1\\
                          & (0, 0, 2)          & (1/8, 1/4, 1/4)&2 \\ 
&& (1/8, 1/2, 1/4)&1 \\
&& (1/4, 3/8, 1/4)&2 \\
&& (3/8, 1/2, 1/4)&1 \\
\hline
$(AC)_4$/$(BC)_4$ (100) superlattice  & (1/2, 1/2, 0) &(0, 1/8, 1/2)  &1\\
                               & (1/2, -1/2, 0) & (0, 3/8, 1/2)  &1\\
                          & (0, 0, 4)          & (1/8, 1/4, 1/2)&2 \\ 
&& (1/8, 1/2, 1/2)&1 \\
& & (1/4, 3/8, 1/2)&2 \\
& & (3/8, 1/2, 1/2)&1 \\
\hline
$A_1B_7C_8$ fcc supercell    & (1, 1, 0) & (0, 0, 1/4) & 1 \\
                          & (1, 0, 1) & (0, 1/4, 1/2) & 3 \\
                          & (0, 1, 1) &  \\
\hline
$A_1B_{15}C_{16}$ bcc supercell & (1,   1,  -1) & (1/8, 1/8, 1/8) & 1 \\
                          & (1,  -1,   1) & (1/8, 5/8, 5/8) & 1\\
                          & (-1,   1,   1) \\
\hline
$A_1B_{31}C_{32}$ simple cubic supercell & (2,   0,   0) & (1/4, 1/4, 1/4)
& 1 \\
                          & (0,   2,   0) \\
                          & (0,   0,   2)  \\
\hline
$A_{8}B_8C_{16}$ SQS16          & (1, -1, 2) &  (0, 1/2, 0)   &1\\
                          & (1, -1 -2) &  (1/2, 0, 0)   &1\\
                          & (1/2, 1/2, 0)&(1/4, 3/4, 1/8) &4\\
                          &             & (0, 1/2, 1/4) &2\\
                          &             & (1/2, 0, 1/4) &2\\
                          &             & (1/4, 3/4, 3/8) &4\\ 
                          &             & (0, 1/2, 1/2) &1\\
                          &             & (1/2, 0, 1/2) &1\\
\end{tabular*}
\end{ruledtabular}
\end{table*}

The {\em enthalpy of mixing} for an alloy $A_{1-x}B_x$ of two components $A$
and $B$ is the difference in energy between the alloy and the
weighted sum of the constituents;
\begin{equation}
\label{eq:deltaH}
\Delta H(x) = E_{A_{p}B_q}-[(1-x)E_A + xE_B]
\end{equation}
where $x=p/(p+q)$.  In the mixing enthalpy calculations, the alloy
lattice constant $a(x)$ is taken as the linear average of the
experimental values of the constituents. The experimental lattice
constants of BAs and GaAs are 4.777 \AA\ and 5.653 \AA, respectively,
whereas the calculated lattice constants are 4.740 \AA\ and
5.615\AA. Any free internal coordinates in the alloy structure were
relaxed using the quantum mechanical forces so that residual forces on
the ions were less than 1 mRyd/bohr.
%

The {\em band gap bowing} parameter $b$ defined by
\begin{equation}
\label{eq:bowing}
E_g(x)=\overline{E}_g(x)-bx(x-1),
\end{equation}
where $\overline{E}_g(x)$ is the weighted linear average of the
individual band gaps of $A$ and $B$. The bowing parameter $b$
represents the deviation of the band gap from this average. Note that
although there are LDA errors in the calculated gaps of the
$A_xB_{1-x}$ alloy as well as the pure constituents A and B, to lowest
order, the LDA error in the bowing $b$ cancels out. 

Considering Eqs.~(\ref{eq:deltaH}) and (\ref{eq:bowing}), we see that
one needs to converge the $k$ representation for a compound $A_pB_q$
as well as for the elemental constituents $A$ and $B$. The standard
way of accomplishing this is to increase the number of $k$ points in
all three systems until convergence is obtained. The disadvantage of
this approach is that it requires {\em absolute} $k$ point convergence
for $A$ and $B$, and separately for $A_pB_q$. A better approach is to
take advantage of {\em relative} $k$ point convergence.\cite{froyen}
The idea is to sample the Brillouin zone {\em equivalently} for $A$,
$B$ and $A_pB_q$. This could be done by considering $A_pA_q$,
$B_pB_q$, and $A_pB_q$ as isostructural solids and sampling the
Brillouin zone of each equally. Then, any relative $k$-point sampling
error cancels out. This is called the {\em method of equivalent $k$
points}.\cite{froyen} In practice, one does not have to calculate
$A_pA_q$ and $B_pB_q$ but instead can calculate $A$ and $B$ in their
primitive unit cells using suitably folded-in $k$ points. Equivalent
$k$ points for the unit cells in this paper are given Table
\ref{kpoints}.

\subsection{Choice of supercells}

\begin{table}[bt] \centering
\caption{All possible pair configurations in the 64 atom cell.}
\label{pairs}
\begin{ruledtabular}
\begin{tabular}{ccc}
Shell number & Lattice vector & Number of  \\
($n$th fcc neighbor)  & [$u,v,w$] (unit $a_0/2$) & equivalent pairs \\
\hline
1 & [0, 1, 1] & 12 \\
2 & [2, 0, 0] & 3  \\
3 & [2, 1, 1] & 12 \\
4 & [2, 2, 0] & 3  \\
6 & [2, 2, 2] & 1  \\
\end{tabular}
\end{ruledtabular}
\end{table}

The calculations of $\Delta H$ [Eq.~(\ref{eq:deltaH})] and $b$
[Eq.~(\ref{eq:bowing})] require supercells. We use B$_1$Ga$_7$As$_8$,
B$_1$Ga$_{15}$As$_{16}$, B$_1$Ga$_{31}$As$_{32}$, and
B$_2$Ga$_{30}$As$_{32}$. The lattice vectors defining the supercells are
given in Table~\ref{kpoints}. The SQS16 supercell is a ``special
quasirandom structure''---a periodic structure with rather small unit
cell whose lattice sites are occupied by $A$ and $B$ atoms so as to
mimic the atom-atom correlation functions of much larger $A_{1-x}B_x$
supercells with random occupations.\cite{sqs}

In the calculations for the band gap bowing of B$_x$Ga$_{1-x}$As
alloys, 64 atom, simple-cubic unit cells were used for both the 3\%
and 6\% boron alloys. In the case of the 3\% alloy, there is one boron
atom in the supercell, but for the 6\% alloy, there are two boron
atoms in the supercell. For this case, the band gap was determined by
taking the weighted average of the gaps for the 5
symmetrically-inequivalent configurations (given in Table~\ref{pairs})
of two boron atoms in the 64 atom supercell. These 5 pairs are the
first through fourth neighbor pairs in an fcc lattice, as well as the
sixth neighbor. (In the 64 atom cell, the fifth fcc neighbor is
equivalent to the first.)  Using the same 64 atom unit cell for the
alloys in the bowing calculations eliminated band gap differences that
can occur due to the k-point folding relations of different
supercells, as discussed by Bellaiche et al. in Ref.~
\onlinecite{bellaiche}.

\subsection{Valence force field model}
\label{secvff}

\begin{table}[tb] \centering
\caption{Parameters used in the VFF model. $d^0_{\text{bond}}$ is the
equilibrium (unstretched) bond length, $\alpha$ is the bond stretch
term, $\beta$ is the bond bending term, and $\sigma$ is the
stretching-bending term. The units of $\alpha$, $\beta$, and $\sigma$
are 10$^3$ dyne.}
\label{vffparams} 
\begin{ruledtabular}
\begin{tabular}{ldddd}
System & d^0_{\text{bond}} (\AA) & \alpha & \beta & \sigma \\ 
\hline
\\
For band gap bowing & & & & \\
\hline
BAs    &   2.0685  & 76.26  & 19.12 & -9.12 \\ 
GaAs   &   2.4480  & 32.15  &  9.37 & -4.10 \\ 
\hline
\\
For mixing enthalpy & & & & \\
\hline
BAs    &   2.0685  & 76.26  &  22.2 & \text{---} \\ 
GaAs   &   2.4480  & 41.19  &  8.94 & \text{---} \\ 
GaN    &   1.9520  & 96.30  &  14.8 & \text{---} \\ 
\end{tabular}
\end{ruledtabular}
\end{table}

In two cases, the total energies or the internal structural parameters
were determined using a classical force field model. This generalized
valence force field model\cite{vff,vff2} (VFF) consists of bond
stretching, bond bending, and bending-stretching springs with force
constants $\sigma$, $\beta$, and $\sigma$, respectively. The force
constants and the equilibrium (no force) bond lengths are given in
Table \ref{vffparams}. The two cases where the VFF model was used are:

(1) The VFF model was used to model the mixing enthalpies of boron
or nitrogen in GaAs in the regime of very low concentrations
(Sec.~\ref{sec:deltaH}). Because well tested force constants exist for
the ``conventional'' VFF model ($\alpha$, $\beta$; no $\sigma$) for
the case of GaAs/GaN, we also used a two-parameter ($\sigma=0$) model
for BAs/GaAs for the sake of consistency with previous calculations.
The parameters for BAs/GaAs were tested by comparing the enthalpies of
formation for structures at higher concentrations calculated using
first principles, as given in Table~\ref{deltaH}.

\begin{table*} \centering
\caption{Enthalpies of formation (LDA and VFF) of ordered (001)
superlattices of BAs/GaAs and the mixing enthalpies of
B$_x$Ga$_{1-x}$As random alloys [Eq.~(\ref{eq:deltaH})]. All the
$\Delta H$ values are positive.}
\label{deltaH}
\begin{ruledtabular}
\begin{tabular}{lcdd}
System & Boron concentration & \multicolumn{1}{c}{$\Delta
H$(meV/atom)} & \multicolumn{1}{c}{$\Delta H$(meV/atom)} \\
 
 & & \multicolumn{1}{c}{(LAPW)} & \multicolumn{1}{c}{(VFF)} \\ 
\hline
\multicolumn{3}{c}{Ordered superlattices:} &\\ 
(BAs)$_1$/(GaAs)$_1$ (001)& 50\% & 186 & 193 \\
(BAs)$_2$/(GaAs)$_2$ (001)& 50\% & 171 & 192 \\
(BAs)$_4$/(GaAs)$_4$ (001)& 50\% & 161 & 191 \\
(BAs)$_\infty$/(GaAs)$_\infty$ (001)& 50\% & 148 &\text{---} \\ 
(BAs)$_1$/(GaAs)$_3$ (001)& 25\% & 132 & 174 \\
\\
\multicolumn{3}{c}{Random alloys:}&\\
BGa$_{7}$As$_{8}$ (fcc)   &   12.5\%  & 73   & 79 \\
BGa$_{15}$As$_{16}$ (bcc) &   6.25\%  & 22   & 30 \\
BGa$_{31}$As$_{32}$ (simple cubic)  &   3.13\%  & 9.5  & 16 \\
\end{tabular}
\end{ruledtabular}
\end{table*}

(2) Full relaxation using first principles' forces of all 64 atom
supercells used in the band gap bowing calculations reported in
Sec.~\ref{sec:bowing} is computionally prohibitive so the
generalized\cite{vff2} VFF model was used to determine internal
coordinates. In one case (B$_1$Ga$_{31}$As$_{32}$), the VFF- and
LAPW-determined coordinates were compared and found to be very
similar. The average difference between the two methods for the Ga--As
bonds was $<.005$ a.u.\ (0.1\%). The largest discrepency was near the
boron site where the VFF predicts a B--As bond length that is $0.01$
a.u.\ (1.5\%) shorter than the LAPW bond length. Using the
VFF-determined internal coordinates of the supercell would be
acceptably accurate only if the band gap is very close to that when
the first-principles' coordinates are used. In the case of the
BGa$_{31}$As$_{32}$ supercell just mentioned, the band gap difference
was only 10 meV, justifying the use of the VFF.

\section{BA\lowercase{s} and the III-A\lowercase{s} family}
\subsection{Expectations from atomic physics}

\begin{figure}[bp]
   \includegraphics[width=.9\linewidth]{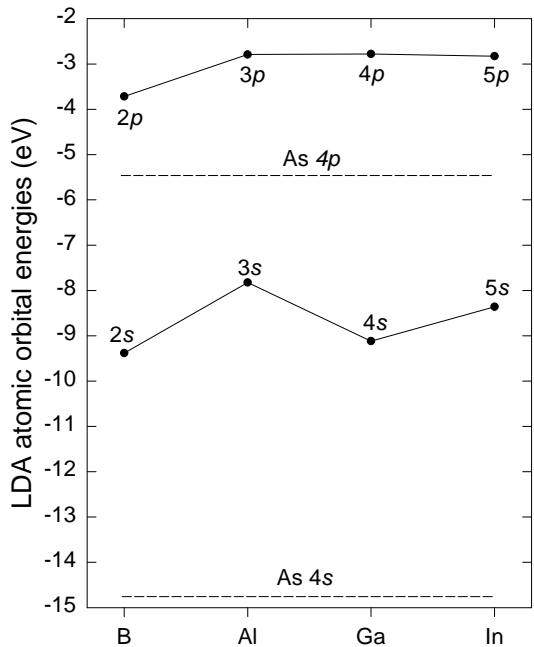}
   \caption{Atomic $s$ and $p$ orbital energies in column III of the
   Periodic Table. Calculated fully relativistically within the local
   density approximation.} 
   
\label{atomicEigenvalues}
\end{figure}

The LDA valence orbital energies for the first four elements of column
III are plotted in Fig.\ \ref{atomicEigenvalues}. The lower value of
the atomic $p$ states of B relative to the atomic $p$ states for Al,
Ga, and In (which are all quite similar) are notable. This lower
value can be understood from the fact that, as a first row element,
the $2p$ states in B need not be orthogonal to lower $p$ states. In
general, the cation $s$ energy should increase going down the column
but there is a kink when moving from Al to Ga due to the introduction
of imperfectly screened $d$ states in Ga. The fact that the $p$ states
in B lie lower in energy, and hence closer to the As $4p$ states, will
lead to a much stronger hybridization of $p$-like states in BAs
relative to the other III-As systems.

\begin{figure}[b] \centering
  \includegraphics[width=.9\linewidth]{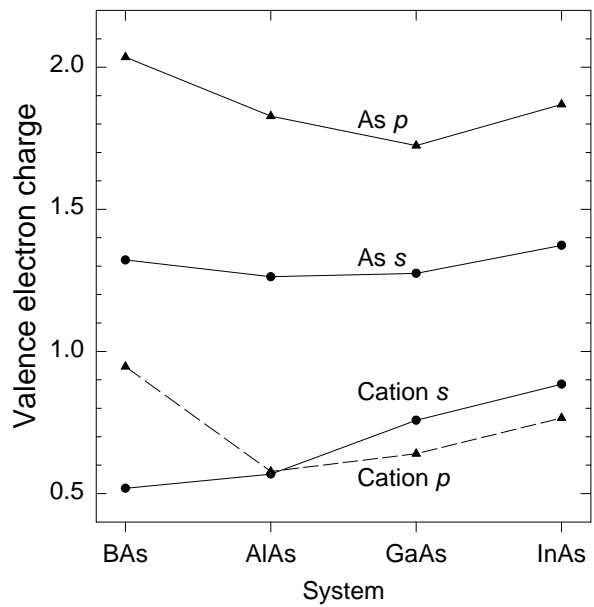}
  \caption{Total valence charge [Eq.~(\ref{eq:VBrho})] enclosed in spheres having the
  tetrahedral radii of Phillips:\cite{phillips} boron 0.853 \AA,
  aluminum 1.230 \AA, gallium and arsenic 1.225 \AA, and indium 1.405
  \AA.}
\label{occ}
\end{figure}

The consequence of this is shown in Fig.\ \ref{occ} which shows the
total valence charge, $q^{\text{tot}}_{\ell}$ of angular momentum
$\ell$ [Eq.\ (\ref{eq:qtot})], enclosed in spheres having the
tetrahedral radii of Phillips.\cite{phillips} It is clear that the
occupation of the B and As $p$-like states in BAs is much higher than
in the other systems (indicating that $p$-$p$ hybridization is much
more pronounced in BAs) and also that there is more charge around the
boron than the other cations. Thus, the distinction between ``cation''
and ``anion'' is not so clear cut in BAs, where both atoms share
similar charge. This is also reflected in the relative ionicities in
the group. According to the Phillips scale\cite{electronegativitynote}
the relative ionicities are .002, .274, .310, and .357 for BAs, AlAs,
GaAs, and InAs, respectively. As a result of the relative ionicities,
the bonding in BAs is almost completely covalent whereas the other
three members of the group have a significant ionic component that
increases in the order AlAs $\rightarrow $ GaAs $ \rightarrow $
InAs. We will see next how these expectations are reflected in the
electronic structure of BAs.

\subsection{Energy bands, band characters, and densities of states in
zinc-blende BA\lowercase{s}}
\label{secbands}

\begin{figure}[tbp]
  \includegraphics[width=\linewidth]{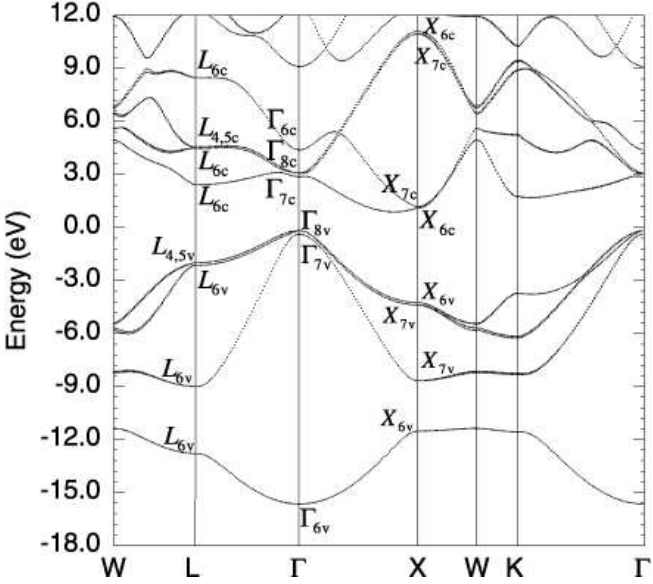}
  \caption{Relativistic energy band structure of BAs. The VBM is taken to
           be zero and the lattice constant is 4.777 \AA.}
   \label{bands}
\end{figure}

\begin{figure}[tbp]
  \includegraphics[width=.7\linewidth]{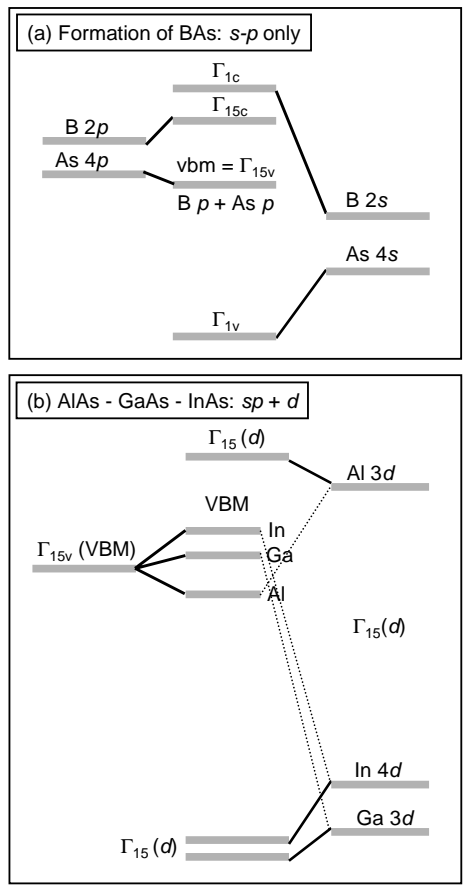}
  \caption{ Schematic diagram of energy levels in the III--As
series. (a) Schematic diagram of energy levels in zinc-blende BAs and
their atomic origins. Because the B-$p$--As-$p$ orbital energy
difference is small relative to other III--V compounds and the
B-$s$--As-$s$ orbital energy difference is large, the $p$-like $\Gamma
_{15\text{c}}$ conduction band state is {\em below} the $s$-like
$\Gamma _{1\text{c}}$ conduction band state. This ordering of the
lower conduction band states is unusual for III--V compounds and is
reminiscent of silicon. (b) Schematic diagram of the effects of
$p$--$d$ coupling on the VBM in AlAs, GaAs, and InAs. The VBM is
driven up by the coupling of the 3$d$(4$d$) states in GaAs(InAs) and
the As 4$p$-like VBM, but in AlAs, this $p$--$d$ repulsion drives the
VBM down.}
\label{schematic}
\end{figure}

The energy bands of zinc-blende BAs are shown in Fig.\ \ref{bands}.
Calculated band energies at high symmetry points are given in Table
\ref{bandeigenvalues}.\cite{perdew,surh} As others have shown
previously,\cite{chu,surh,stukel,ferhat} the band gap of BAs is
indirect; the CBM is along the $\Delta $ line between the $\Gamma $
and $X$ points---at approximately 0.82$(1,0,0)2\pi /a$. An unusual
feature of the band structure is the character of the CBM at
$\Gamma$. In BAs, the CBM is the $p$-like $\Gamma _{7{\text{c}}}$
state ($\Gamma _{15\text{c}}$ if spin-orbit interaction not
included). Only the semiconductors silicon and BP share this
feature. In most semiconductors, the lowest state at $\Gamma$ is the
singly degenerate
$s$-like state. The origins of this feature of the BAs band structure
can be understood in the context of the tight-binding model of
Harrison.\cite{harrison} According to this model, the bonding ($\Gamma
_{15\text{v}}$) and antibonding ($\Gamma _{15\text{c}}$) $p$ states at
$\Gamma$ are given by
\begin{equation}
E(\Gamma _{15})=\frac{\varepsilon^c_p+\varepsilon^a_p}{2}\pm
\sqrt{\left( \frac{\varepsilon^c_p-\varepsilon^a_p}{2} \right) ^2
+ \left(4E_{pp}\right)^2}
\label{peq}
\end{equation}
where $\varepsilon^c_p$ and $\varepsilon^a_p$ are the $p$ atomic
orbital energies for the cation (c) and the anion (a) and the
interatomic term $E_{pp}$ is proportional to $1/d^2_{\text{bond}}$
where $d_{\text{bond}}$ is the bond length. The bonding ($\Gamma
_{1\text{v}}$) and antibonding ($\Gamma _{1\text{c}}$) $s$ states at
$\Gamma$ are given by
\begin{equation}
E(\Gamma _{1})=\frac{\varepsilon^c_s+\varepsilon^a_s}{2}\pm
\sqrt{\left( \frac{\varepsilon^c_s-\varepsilon^a_s}{2} \right) ^2
+ \left(4E_{ss}\right)^2}
\label{seq}
\end{equation} 
where $\varepsilon^c_s$ and $\varepsilon^a_s$ are the $s$ atomic
orbital energies for the cation (c) and the anion (a) and the
interatomic term $E_{ss}$ is again proportional to
$1/d^2_{\text{bond}}$. A schematic diagram of the BAs energy levels in
this model is shown in Fig.~\ref{schematic}a.

\begin{table}[tbp] \centering
\caption{Comparison of BAs band energies for LDA and GW.\cite{surh}
Two LDA values are given: all-electron LAPW values from this work and
pseudopotential planewave (PP) values.\cite{surh} All the calculations
used the experimental lattice constant of 4.777 \AA. The
exchange-correlation potential used in the PP calculations was not
reported. For our LAPW calculations, the exchange-correlation of
Perdew and Wang\cite{perdew} was used (See Sec.~\ref{details} for
details.)}
\label{bandeigenvalues}
\begin{ruledtabular}
\begin{tabular}{lccc}
   & \multicolumn{1}{c}{LDA-PP (eV)} &
   \multicolumn{1}{c}{LDA-LAPW (eV)} & \multicolumn{1}{c}{GW (eV)} \\
\hline 
$\Gamma _{6c}$ & 4.5 & 4.57 & 5.5 \\ 
$\Gamma _{8c}$   $(\times 2)$ & 3.3 & 3.26 & 4.2 \\
$\Gamma _{7c}$ & 3.1 & 3.05 &   4.0 \\
$\Gamma _{8v}\,(\times 2)$ & 0.00 & 0.00 & 0.00 \\ 
$\Gamma _{7v}$ & -0.22 & -0.21 & -0.22 \\ 
$\Gamma _{6v}$ & -15.5 & -15.46 & -16.7 \\ 
\hline $X_{6c}$ & 11.4 & 11.27 & 13.1 \\
   $X_{7c}$ & 11.2 & 11.13 & 12.9 \\ 
$X_{6c}$ & 1.38 & 1.274 & 1.86 \\
   $X_{7c}$ & 1.36 & 1.359 & 1.93 \\ 
$X_{6v}$ & -4.1 & -4.07 & -4.5 \\
   $X_{7v}$ & -4.2 & -4.21 & -4.6 \\ 
$X_{7v}$ & -8.6 & -8.49 & -9.5 \\
   $X_{6v}$ & -11.3 & -11.35 & -12.2 \\ 
\hline $L_{6c}$ &   8.8 & 8.65 & 9.8 \\
 $L_{4,5c}$ & 4.8 & 4.74 & 5.7 \\
 $L_{6c}$ & 4.7
   & 4.66 & 5.6 \\
 $L_{6c}$ & 2.6 & 2.60 & 3.3 \\ 
$L_{4,5v}$ & -1.8 &
   -1.81 & -2.0 \\ 
$L_{6v}$ & -1.9 & -1.96 & -2.1 \\
 $L_{6v}$ & -8.8 &
   -8.82 & -9.7 \\ 
$L_{6v}$ & -12.6 & -12.62 & -13.6 \\ 
\end{tabular}
\end{ruledtabular}
\end{table}

The reversal of the conduction band states in the BAs band structure
relative to other III--V compounds is a result of two effects: (i) a
small $\Gamma _{15\text{v}}$--$\Gamma _{15\text{c}}$
bonding/antibonding repulsion due to the small $E_{pp}$, as described
by Eq.~(\ref{peq}), and (ii) a large $\Gamma _{1\text{v}}$--$\Gamma
_{1\text{c}}$ repulsion of due to the small
$(\varepsilon^B_s-\varepsilon^{As}_s)^{2}$, as described by
Eq.~(\ref{seq}). As is evident from Fig.~\ref{schematic}, a small
$\Gamma _{15\text{v}}$--$\Gamma _{15\text{c}}$ bonding/antibonding
repulsion will lower the $\Gamma _{15\text{c}}$ state and a large
$\Gamma _{1\text{v}}$--$\Gamma _{1\text{c}}$ bonding/antibonding
repulsion will raise the $\Gamma _{1\text{c}}$. In BAs these two
effects are enough to reverse the ordering of the $\Gamma
_{15\text{c}}$ and $\Gamma _{1\text{c}}$ states.

The large $\Gamma _{1\text{v}}$--$\Gamma _{1\text{c}}$
bonding/antibonding difference is due to the large $E_{ss}$ due to the
short bond length in BAs. Empirical tight binding has also shown that
the $\Gamma_{1\text{v}}$--$\Gamma_{1\text{c}}$ repulsion is much more
pronounced in BN, BP, and BAs than in other III--Vs.\cite{ferhat} The
small $\Gamma _{15\text{v}}$--$\Gamma _{15\text{c}}$
bonding/antibonding repulsion is due to the small differences in the
$p$ atomic orbitals of the boron and arsenic, as shown in
Fig.~\ref{atomicEigenvalues}. That is, the first term under the
radical in Eq.~(\ref{peq}) is small relative to other III--Vs.

\begin{figure}[tbp]
\includegraphics[width=\linewidth]{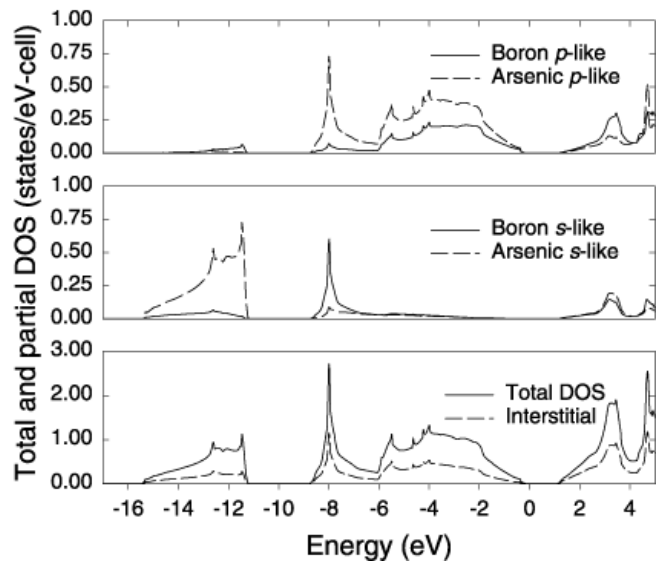}
\caption{Partial and total densities of states for BAs
[Eq.~(\ref{eq:dos})] inside muffin tin spheres with radii 0.853 \AA\ for
boron and 1.225 \AA\ for arsenic.\cite{phillips} Note the strong
mixing of B and As $p$ states. Figure \ref{occ} shows a quantitative
measure of the $p$-$p$ mixing in the each of the III--As compounds.}
\label{dos}
\end{figure}
\begin{figure*}[tbp]
\centerline{\includegraphics[width=.9\linewidth]{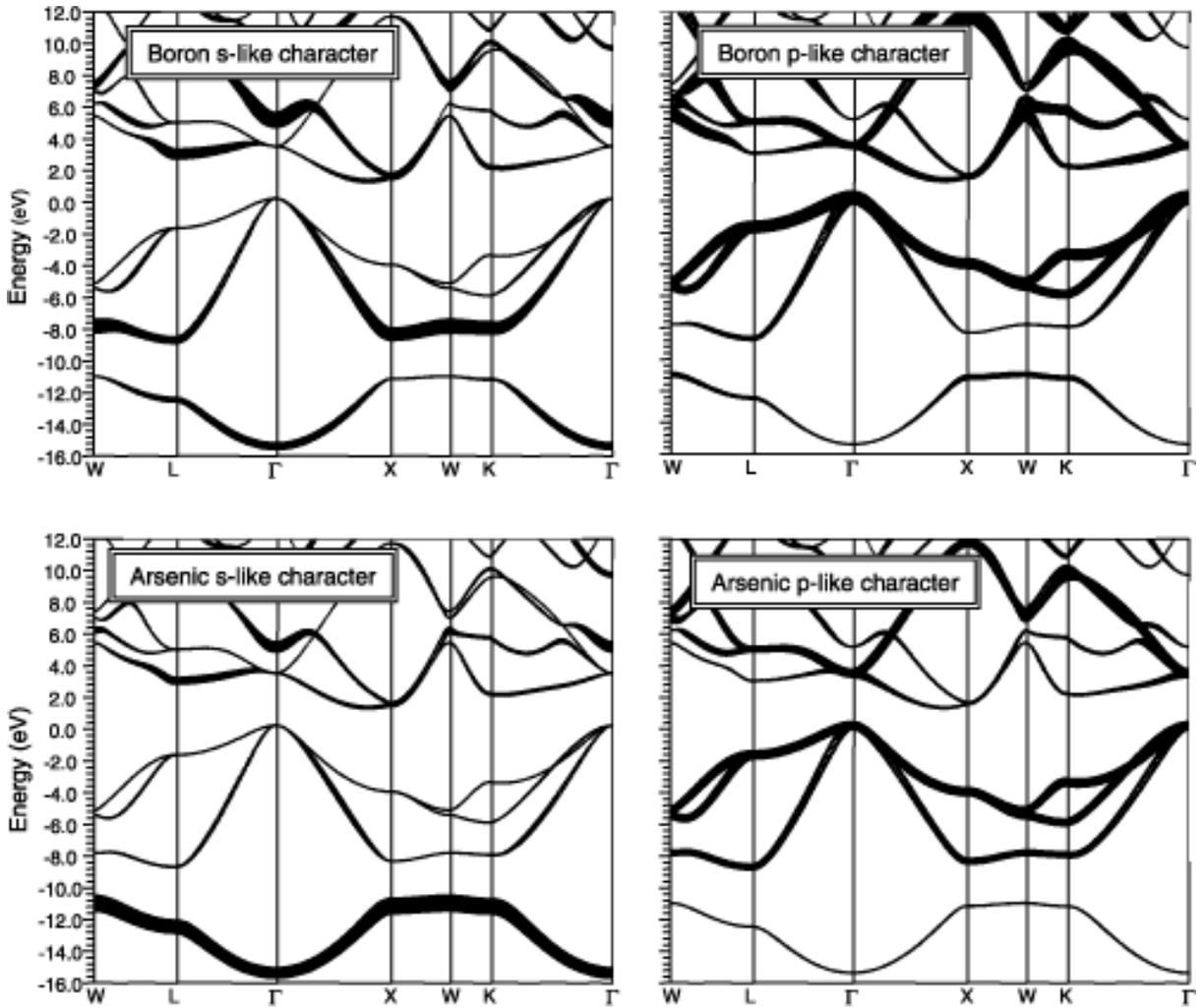}}
\caption{Band characters of BAs [Eq.~(\ref{ql})] plotted as a function
of $k$. The thickness of the lines denotes a relative amount of a
given character. The B and As $p$ states are stongly mixed at both the
VBM and the CBM whereas the As $s$ band is distinct. There is some
mixing of As $p$ and B $s$ states in the bottom of the $p$-$p$
hybridized bands.}
\label{bandcharacters}
\end{figure*}

The strong hybridization of the B and As $p$ states is visible in the
partial densities of states plot [Eq.~(\ref{eq:dos})] in Fig.\
\ref{dos} and band character plots in Fig.~\ref{bandcharacters}
[Eq.~(\ref{ql})] where the width of the lines indicates the
amount of the character indicated in the plot label. From these plots,
one can see that the $s$ band of arsenic between $-12$ and $-16$ eV is
distinct and has very little mixing with other states. This is similar
to the other three III--As systems, although in the case of BAs the
arsenic $s$ band is wider and somewhat lower in energy. Two unique
features of the BAs DOS compared to other III--V semiconductors is the
pronounced boron $p$ character at the VBM and anion $p$ character at
the CBM. The stronger $p$-$p$ hybridization in BAs relative to the
rest of the III-As is due to the proximity of the orbital energies of
B and As and the short bond length in the compound.

\subsection{Trends in band gap and inter-valley
energy differences along the III-As family} 

\begin{figure}[tbp]
\includegraphics[width=\linewidth]{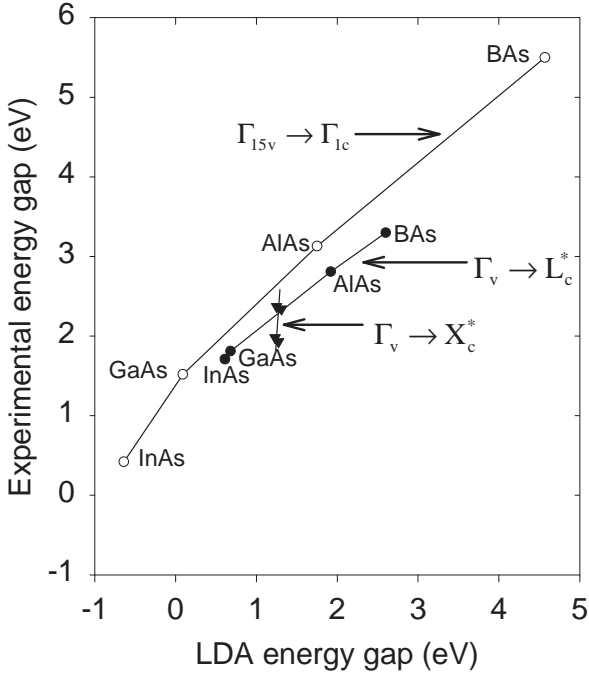}
\caption{Trends in the LDA vs. experimental band gaps of the III--As
family. For the $\Gamma _{15\text{v}}\rightarrow \Gamma _{1\text{c}}$
and the $\Gamma_{15\text{v}}\rightarrow L_{\text{c}}$ gaps, the errors
are nearly constant. The LDA errors are $\sim$1--1.5 eV and
$\sim$\,0.7--1.1 eV for the $\Gamma _{15\text{v}}\rightarrow \Gamma
_{1\text{c}}$ and the $\Gamma_{15\text{v}}\rightarrow L_{\text{c}}$
gaps, respectively.}
\label{gaps}
\end{figure}

Figure \ref{gaps} shows the LDA-calculated vs. experimental band gaps
for the III--As family, MAs (M = B, Al, Ga, In). For all but BAs, LDA
energy band gaps are taken from the LAPW calculations reported in
Ref.~\onlinecite{wei}. Experimental band gaps are from
Ref.~\onlinecite{semicon} For BAs, for which no reliable experimental
data exists, we used the $GW$ band gap from Ref.\
\onlinecite{surh}. Because of the unusual ordering of the conduction
bands in BAs, the direct gap is not the $\Gamma
_{15\text{v}}\rightarrow \Gamma _{1\text{c}}$ gap but the $\Gamma
_{15\text{v}}\rightarrow \Gamma _{15\text{c}}$. The figures show the
$\Gamma _{15\text{v}}\rightarrow \Gamma _{1\text{c}}$ gaps (which are
the direct gaps in AlAs, GaAs, and InAs). We find in Fig.\ \ref{gaps}
that the LDA vs. expt. $\Gamma _{15\text{v}}\rightarrow \Gamma
_{1\text{c}}$ and the $\Gamma_{15\text{v}}\rightarrow L_{\text{c}}$
gaps lie almost on a straight line. The LDA error of these gaps in the
III--As family is approximately constant, even for BAs. For the
$\Gamma _{15\text{v}}\rightarrow \Gamma _{1\text{c}}$ gaps, the
magnitude of the error is $\sim$1--1.5 eV. For the $\Gamma
_{15\text{v}}\rightarrow L_{\text{c}}$ gaps, the magnitude of the
error is $\sim$\,0.7--1.1 eV. The experimental $\Gamma
_{15\text{v}}\rightarrow X_{\text{c}}$ gaps only vary by $\sim$\,0.5 eV
through the series and corresponding LDA gaps are nearly identical,
resulting in a tight clustering of the $\Gamma
_{15\text{v}}\rightarrow X_{\text{c}}$ points in Fig.\ \ref{gaps}.

\begin{figure}[tbp]
\includegraphics[width=.95\linewidth]{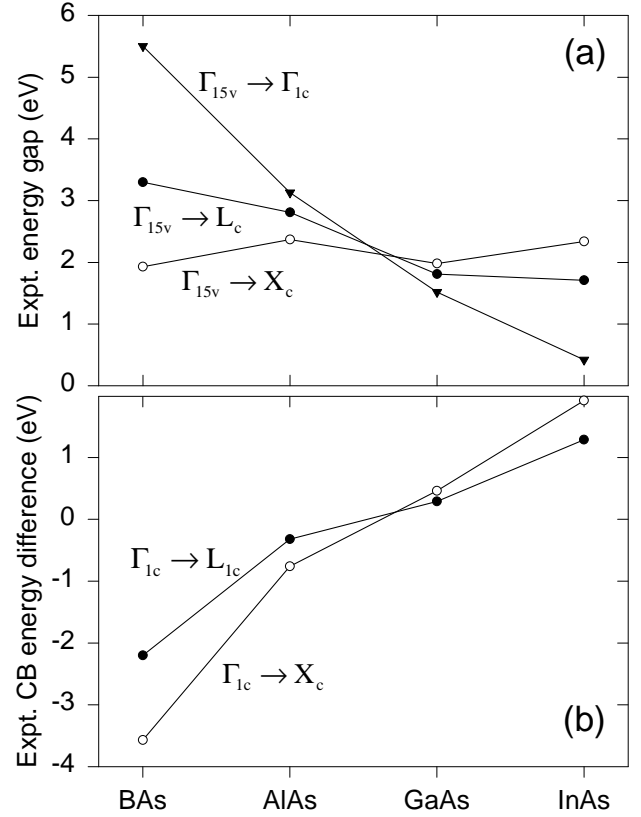}
\caption{Trends in the inter-valley energy seperations of the III--As
family. Note the crossing of the $\Gamma_{15\text{v}}\rightarrow
X_{\text{c}}$ gaps and the $\Gamma_{15\text{v}}\rightarrow
\Gamma_{1\text{c}}$ gaps between AlAs and GaAs. Thus, GaAs and InAs
are direct gap materials whereas BAs and AlAs are indirect $X$ gap
materials.}
\label{intervalley}
\end{figure}

In Fig.\ \ref{intervalley}(a), we show the $\Gamma
_{15\text{v}}\rightarrow \Gamma _{1\text{c}}$,
$\Gamma_{15\text{v}}\rightarrow L_{\text{c}}$, and $\Gamma
_{15\text{v}}\rightarrow X_{\text{c}}$ gaps for the III--As family.
We see: (i) the ordering of the conduction
band states in BAs and AlAs is
$X_{1\text{c}}<L_{1\text{c}}<\Gamma_{1\text{c}}$ whereas (ii) the
ordering is $\Gamma_{1\text{c}}<L_{1\text{c}}<X_{1\text{c}}$ for GaAs
and InAs, (iii) the $X_{1\text{c}}$ and $L_{1\text{c}}$ states are
closely spaced in AlAs and GaAs, (iv) BAs is ``strongly indirect;''
$L_{1\text{c}}$ and $\Gamma_{1\text{c}}$ are far above the CBM (which
is close to $X_{1\text{c}}$), and (v) a crossing of the lowest $X$ and
$L$ conduction states occurs between AlAs and GaAs, but the $\Gamma
_{15\text{v}}\rightarrow \Gamma _{1\text{c}}$ gaps decrease so rapidly
in the series that the $\Gamma_{1\text{c}}$ states lie below the
$L_{1\text{c}}$ in GaAs and InAs, and thus these two members of the
family are direct band gap materials (whereas BAs and AlAs are
indirect $X$ gap materials).

Figure \ref{intervalley}(b) shows the conduction band intervalley
differences $\Gamma_{1\text{c}}\rightarrow X_{\text{c}}$ and
$\Gamma_{1\text{c}}\rightarrow L_{1\text{c}}$. Both differences show a
smoothly decreasing trend from AlAs to GaAs to InAs, but an
anomalously large decrease from BAs to AlAs. Again, this is a result
of the enhanced $s$--$s$ repulsion in BAs. Because the
$\Gamma_{1\text{c}}$ state has much more $s$ character than either the
lowest $X$ and $L$ conduction states (see Fig.\ \ref{bandcharacters}),
the effect of the $s$--$s$ repulsion is greater for the
$\Gamma_{1\text{c}}$ than for these other two states, driving it up
relative to them. This effect is more pronounced for the
$\Gamma_{1\text{c}}\rightarrow X_{\text{c}}$ difference because the
lowest $X$ conduction state has less $s$ character than the lowest $L$
conduction state.

\subsection{Charge density and ionicity trends in the III-As family}
\label{secrho}

\begin{figure}[tbp]
\includegraphics[width=\linewidth]{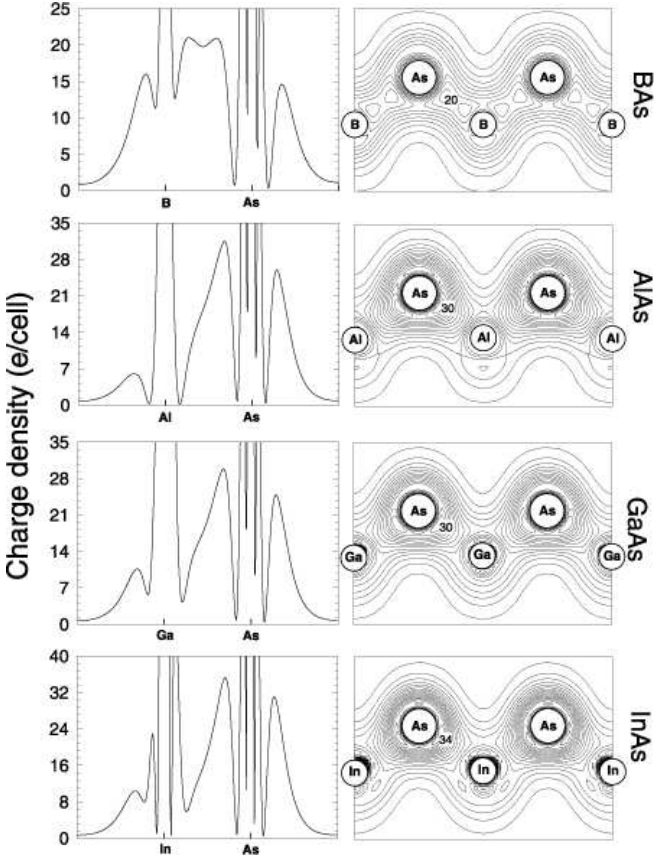}
\caption{Valence charge densities $\rho_{\text{val}}$
[Eq.~(\ref{eq:VBrho})] for the III-As family, excluding the cation and
anion $d$ states. The contour plots use the same scale as the
corresponding line plots. The contour spacing is 2 e/cell. The range
of the line plots is from one tetrahedral interstitial site, through
the bond, and to the neighboring tetrahedral interstitial site. Note
the striking asymmetry of the bonding charge for the three systems
AlAs, GaAs, and InAs, while the bond charge in BAs reveals a covalent,
double-hump feature similar to that found in diamond C, Si, or Ge.}
\label{rho}
\end{figure}

The valence charge density [upper four bands, $d$ bands not included;
Eq.~(\ref{eq:VBrho})] in the (110) plane is shown for the four members
of the III-As family in Fig.\ \ref{rho}. We see that the charge
densities of AlAs, GaAs, and InAs are drawn towards the anion and
exhibit a ``single hump'' in the bond charge. (The relative shifts in
the bond charge, $d_{\text{max}}/d_{\text{bond}}$, are $\sim$\,0.68 for
AlAs and GaAs, and $\sim$\,0.71 for InAs.)  The relative shift of charge
towards the anion (As) is due to the difference in electronegativity
of the anion and the cation and reflects the partially ionic nature of
the bond. In contrast, because of the almost entirely covalent nature
of the bonding in the BAs system, in this system the charge density
features a ``double hump'' bond similar to the bond charges in diamond
C and Si.\cite{wang,lu} A similar analysis of the charge density of
BAs was reported earlier by Wentzcovitch and
Cohen.\cite{wentzcovitch} We see that BAs is unique in the III-As
family in that the bonding is very covalent. This is evidenced by the
exceptionally strong hybridization of $p$ states from the anion and
cation (see Figs. \ref{occ}, \ref{dos}, and \ref{bandcharacters}) and
the significant bonding charge visible in the valence charge density
which is similar to the bonding charge in diamond C and Si.

\begin{figure*}[tbp]
\includegraphics[width=.9\linewidth]{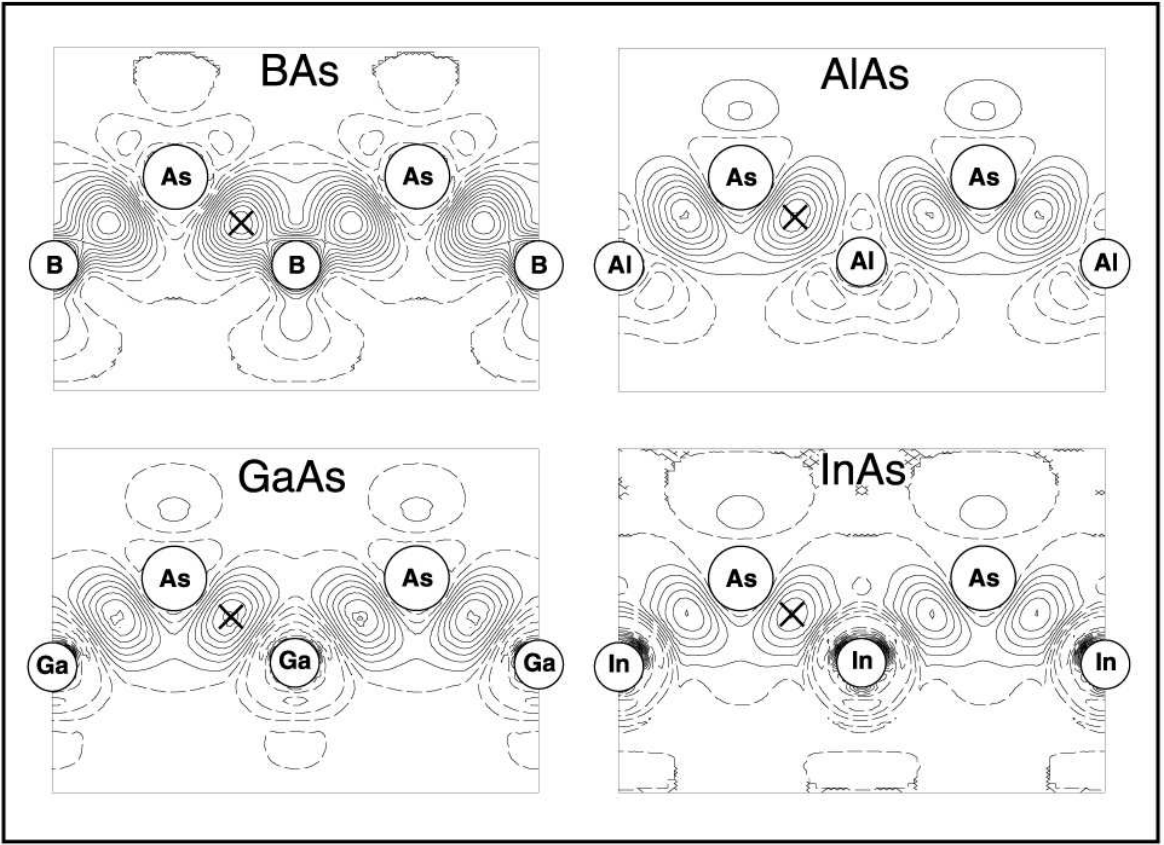}
\caption{Deformation density $\Delta\rho_{{\rm {val}}}$
($\rho^{\text{solid}}_{\text{val}}-\rho^{\text{atom}}_{\text{val}}$)
plots for the III-As family in the (110) plane. Contours marked with
dashed lines correspond to negative values of $\Delta\rho_{{\rm
{val}}}$. The units are e/\AA$^3$\ and the contour spacing is
0.02. The maxima are marked by crosses. To determine $\Delta\rho_{{\rm
{val}}}$, the overlapping charge densities of free atoms are
subtracted from the self-consistent charge density of the solid. The
atomic configurations for the free atoms are $d^{10}s^2p^1$ and
$s^2p^3$ for the cations and anions, respectively. The $d$ bands of
the cations, but not of the anions, are included in the valence
bands.}
\label{deltarho}
\end{figure*}

It is interesting to note that, in contrast to AlAs, GaAs, and InAs
where charge is drawn towards the anion, the nearly symmetric bond
charge distribution shown in the valence charge plot of BAs is not
centered exactly around the bond center but is actually drawn slightly
toward the {\em cation} B. (The distance between the bond center and
the boron position is about 45\% of the total bond length.) This would
indicate that whatever small ionic component exists in BAs results not
from charge transfer towards the {\em anion} but from charge transfer
towards the {\em cation} B. Further evidence of an ionic component in
BAs where the boron atom acts as the anion is shown in the valence
charge difference
plots [Eq.~(\ref{eq:rhodiff})] in Fig. \ref{deltarho}. In AlAs, GaAs,
and InAs, the maximum (marked as a cross) in $\Delta \rho
_{\text{val}}$ occurs closer to the As anion than to the cation,
consistent with the reported ionicities. However, in BAs, the maximum
in $\Delta \rho _{\text{val}}$ occurs closer to the B atom. In
addition, the positive region around the B atom in $\Delta \rho
_{\text{val}}$, which does not occur for the other members of the
III--As family, is further indication of a slight ionic component in
BAs where B plays the role of the {\em anion} rather than the
cation. This reversal of the cation and anion roles in BAs (and in BP)
is predicted by the Phillips scale of
electronegativities\cite{electronegativitynote} and has been discussed
previously by Wentzcovitch and Cohen.\cite{wentzcovitch}

\section{Band offsets:
BA\lowercase{s}/G\lowercase{a}A\lowercase{s}/A\lowercase{l}A\lowercase{s}}
\label{offsetsec}

\begin{figure}[tbph]
\centerline{\includegraphics[width=.7\linewidth]{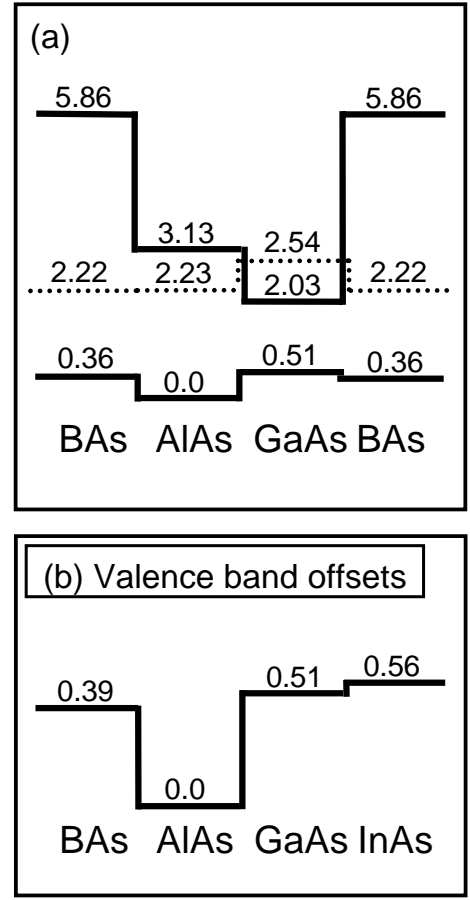}}
\caption{Valence and conduction band offsets (in eV) for BAs/GaAs/AlAs
and for the III-As family. The valence band offsets are calculated
directly using Eq.~(\ref{vboeqn}). The conduction band offsets are
obtained by adding the experimental band gap values to the valence
band offsets. Part (a) shows the both the valence band and conduction
band offsets for BAs/AlAs, AlAs/GaAs, and GaAs/BAs. Both the direct
$\Gamma _{15\text{c}}$ gap (solid line) and the $X_{1\text{c}}$ gap
(dotted line) are shown. The height of the BAs VBM was determined by
averaging the valence band offsets of BAs with respect to AlAs and
GaAs. Part (b) shows the valence band offsets in the III--As
family. Surprisingly, the VBM of BAs is found to be above that of AlAs
and not very far below that of GaAs.}
\label{offsetplot}
\end{figure}

The offset $\Delta E_v(AX/BY)$ between the valence band maxima of two
semiconductor compounds $AX$ and $BY$ forming a heterostructure is one
of the most important parameters in interfacial structures deciding
both transport and quantum confinement. The results for the natural
(unstrained) band offsets [Eq.~(\ref{vboeqn})] for BAs/GaAs and
BAs/AlAs are plotted in Fig.\ \ref{offsetplot}. The {\em conduction}
band offsets shown in the figure are determined by adding the
experimental gaps to the calculated valence band offsets. The natural
band offset for AlAs/GaAs has been computed previously\cite{shw-apl}
and has been included here to contrast the offset behavior of
``typical'' heterojunctions and those of heterojunctions with BAs.

The ordering of the VBM of BAs in the III-As is somewhat
surprising. Because the VBM decreases systematically in the order InAs
$\rightarrow$ GaAs $\rightarrow$ AlAs [Fig.\ \ref{offsetplot}(b)], one
might expect that the VBM of BAs would lie {\em below} AlAs, but in
fact, the VBM of BAs is above AlAs, nearly as high as GaAs. The
ordering of the VBMs for InAs, GaAs, and AlAs can be understood
qualitatively using a tight-binding argument analogous to
Eq.~(\ref{peq}) for $\Gamma_{15\text{c}}$ when a $p$-$d$ term is
added.
The effects of this additional term are
indicated schematically in Fig.\ \ref{schematic}b. The ordering of
the VBMs is as follows:

{\em I\lowercase{n}A\lowercase{s} relative to
G\lowercase{a}A\lowercase{s}:} Two effects account for the slightly
higher VBM of InAs relative to that of GaAs.\cite {shw} (i) InAs has a
longer bond length than GaAs. This reduces $E_{pp}$ of Eq.~(\ref{peq})
and tends to drive the $\Gamma _{15\text{v}}$ VBM up relative to
GaAs. (ii) In tetrahedral symmetry, the cation $d$ and anion $p$
states share the same $\Gamma_{15}$ symmetry and hence can interact
through the potential matrix element
$\langle\phi_{cation}^d|V|\phi_{anion}^p\rangle=E_{dp}$. The mostly
anion $p$-like VBM is repelled upwards by the Ga/In $d$ states by an
amount $E^2_{dp}/(\varepsilon^{anion}_p-\varepsilon^{cation}_d)$. This
$p$-$d$ repulsion\cite{shw,shw-apl} is slightly stronger in InAs where
the cation $4d$ states are shallower and less localized than the Ga
$3d$ state, as shown schematically in Fig.\ \ref{schematic}b.

{\em G\lowercase{a}A\lowercase{s} relative to
A\lowercase{l}A\lowercase{s}:} The difference between the VBM of GaAs
and AlAs can also be understood in terms of this $p$--$d$ repulsion
effect. In GaAs, the deep gallium $d$ states couple to the arsenic $p$
states, driving the VBM up, but in AlAs, there are no aluminum $d$
states below the VBM. Instead, the interaction of the arsenic $p$
states at the VBM and the {\em unoccupied} aluminum $d$ states {\em
above} the VBM drives the AlAs VBM {\em down}. Thus, $p$--$d$
repulsion has the opposite effect on the VBM in GaAs and AlAs (Fig.\
\ref{schematic}b).

{\em A\lowercase{l}A\lowercase{s} relative to BA\lowercase{s}:} Two
effects increase the VBM of BAs relative to the VBM of AlAs. (i)
Because the unoccupied $d$ states in BAs lie very high in energy
relative to the VBM, the $p$--$d$ repulsion effect that drives the VBM
down in AlAs is weaker in BAs. (ii) The more important effect is the
unusual character of the VBM in BAs. In the rest of the III-As family,
the character of the VBM is primarily anion $p$-like, but in BAs the
bonding is much more covalent and the VBM comes from both the anion
and the cation (Fig.\ \ref{bandcharacters}). Since the cation $p$
levels are higher in energy than the As $p$ levels (Fig.\
\ref{atomicEigenvalues}), any admixture of cation $p$ character pulls
the VBM up. These two effects, lack of $p$--$d$ repulsion and strong
hybridization of the cation $p$ character into the VBM, account for
the high VBM of BAs relative to AlAs.

{\em G\lowercase{a}A\lowercase{s} relative to BA\lowercase{s}:} While
the VBM of BAs and GaAs both lie above that of AlAs for the reasons
given above, the VBM of BAs lies only slightly below that of GaAs
despite the much smaller lattice constant ($\sim$17\% mismatch) and
the lack of $p$--$d$ repulsion. This is due to the unusual cation $p$
hybridization into the VBM of BAs which raises the VBM by several
tenths of an eV.

Finally, we note that the transitivity among the band offsets of
BAs/AlAs/GaAs is similar to other compounds where the lattice
mismatches are not as large. By transitivity we mean that the band
offset $\Delta E_v(A/C)$ between two compounds $A$ and $C$ is well
approximated as the sum of the band offsets between compounds $A$ and $B$
and between compounds $B$ and $C$, i.e.,
\begin{equation}
\Delta E_v(GaAs/BAs)\cong
\Delta E_v(BAs/AlAs)+\Delta E_v(AlAs/GaAs).
\end{equation}
In this case, $\Delta E_v(GaAs/BAs)$ calculated directly yields 0.19
eV but $\Delta E_v(BAs/AlAs)+\Delta E_v(AlAs/GaAs)$ yields 0.12 eV, a
non-transitivity difference of 0.07 eV.

\section{BA\lowercase{s}-G\lowercase{a}A\lowercase{s} alloys}
\label{secalloy}
\subsection{Bond lengths and bond angles in the alloys}

Interest in BAs--GaAs alloys centers around the hope that boron will
modify the GaAs band gap similarly to nitrogen without the adverse
effect\cite{geisz2} of introducing localized states that reduce
carrier diffusion length. We know from the theory of III--V
alloys\cite{shw-bowing,magri} that the optical properties are decided
by both the atomic relaxation and by charge transfer.  Thus, we first
study the bond relaxation around a boron substitutional impurity.

\begin{figure}[tb]
\includegraphics[width=\linewidth]{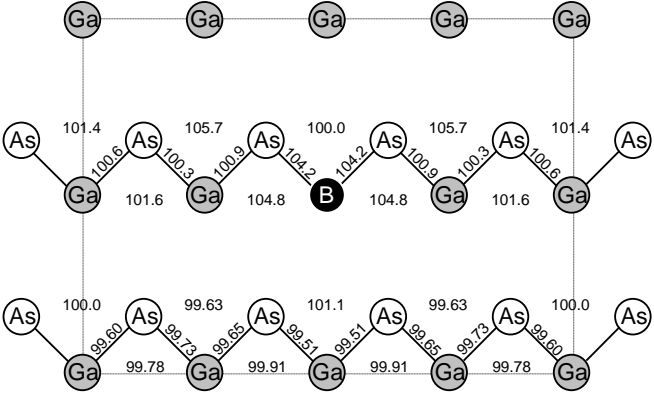}
\caption{Bond lengths and angles in BGa$_{31}$As$_{32}$ in the (110)
plane. Bond lengths are shown as a percentage of the ideal GaAs bond
length, except for the B--As bonds where the values indicate a
percentage of the ideal BAs bond length. The numbers that lie between
a triplet of atoms indicate the bond angle between the three atoms as
a percentage of the ideal bond angle in the zinc-blende structure of
$109.47^\circ$. The rectangle represents the supercell boundary.}
\label{bondlengths}
\end{figure}

The bond lengths and angles in the (110) plane of GaAs with $\sim$3\%
boron substitution (B$_1$Ga$_{31}$As$_{32}$ supercell) are shown in
Fig.\ \ref{bondlengths}. Bond lengths are shown as a percentage of the
pure bulk GaAs bond length, except for the B--As bonds where the
values indicate a percentage of the pure bulk BAs bond length. The
numbers that lie between a triplet of atoms indicate the bond angle
between the three atoms as a percentage of the ideal bond angle in the
zinc-blende structure of $109.47^\circ$. The rectangle represents the
supercell boundary. We note several observations:

(i) Bond angles near boron increase by up to 5\% from their ideal
value. This helps accomodate the Ga--As and B--As alloy bonds
(hereafter $R_{GaAs}$ and $R_{BAs}$, respectively) to keep the bond
lengths close to their ideal values, $R_{BAs}^0$ and
$R_{GaAs}^0$ in the pure binary compounds.

(ii) The B--As bond decreases from the GaAs value towards the BAs
value and ends up only 4\% higher than the value for pure BAs.

(iii) The average bond length relaxation parameter\cite{martins}
\begin{equation}
\epsilon=[R_{GaAs}(x)-R_{BAs}(x)]/[R^0_{GaAs}-R^0_{BAs}] 
\end{equation}
is 0.76. ($\epsilon=0$ when there is no relaxation and 1 when the
relaxation is full.) Thus, assuming $R_{BAs}$ to be Vegard-like (as in
VCA) overestimates the bond length by $\sim$13\%.


We also modeled a 50\%-50\% random alloy using a 32-atom special
quasirandom structure (SQS16 in Table~\ref{kpoints}).\cite{sqs} 
We found that the distribution of the bond lengths
has the expected bimodal form for a random binary
alloy,\cite{bimodal} and the B--As bonds
are
generally larger than the ideal B--As bond length while the Ga--As
bond lengths are smaller than the ideal Ga--As bond length.

\subsection{Bowing in the dilute alloy}
\label{sec:bowing}
Both measurements and calculations indicate that, for isovalent
semiconductor alloys, the deviation of the band gap, $\Delta E_g(x)$,
from the average band gap, $\overline{E}_g(x)$, of the constituents
is often well described by a quadratic
term,\cite{shw-bowing,magri,ld-bowing,bernard}
$$
\Delta E_g(x)=bx(x-1).
$$
For typical semiconductor alloys, the bowing parameter $b$ is normally
less than 1 eV and is independent\cite{semicon} of the concentration
$x$. However, in GaAs$_{1-x}$N$_x$ alloys, where the lattice mismatch
is large and the bond strength differences are significant, the bowing
parameter is strongly composition dependent\cite{bellaiche2} and can
be as large as 20 eV.  Due to the large lattice mismatch of BAs and
GaAs and the strong B--As bonds, one might expect that
B$_x$Ga$_{1-x}$As alloys would also show such large and
composition-dependent bowing.  We calculated the
$\Gamma_{15\text{v}}\rightarrow\Gamma_{1\text{c}}$ gaps in the
experimentally-relevant composition range of 0--10\%. The bowing was
calculated using 64 atom supercells, as described in
Sec. \ref{details}. In this composition range, the direct
$\Gamma_{15\text{v}}\rightarrow\Gamma_{1\text{c}}$ gap is the smallest
gap. The bowing was calculated using the LDA
$\Gamma_{15\text{v}}\rightarrow\Gamma_{1\text{c}}$ gaps of BAs, GaAs,
(4.64 and 0.28 eV, respectively) and the LDA gaps at two alloy
compositions, 3\% and 6\%.  At the 3\% composition, the bowing is
$b=3.4$ eV and at 6\% the bowing is $b=3.6$ eV. Thus, the band gap
bowing is $\sim$3.5 eV and relatively composition-independent in this
composition range. A recent experimental study,\cite{geisz} estimated
the bowing to be $b=1.6\pm 0.3$ or $2.3\pm 0.3$ using {\em theoretical
estimates} for the direct gap of cubic BAs of 3.56\cite{stukel} and
4.23.\cite{prasad} However, because of the LDA error in the band gaps
(see Fig.~\ref{gaps}), these estimated band gaps are too small, and
thus, the bowing is underestimated. Because the LDA errors for the
band gaps are approximately constant for the III--As series, the
bowing can be reliably calculated if {\em all} the points ($E_A$,
$E_B$, and $E_{A_{1-x}B_x}$) are taken consistently from LDA
calculations (as we did) because only the relative values are
important. For computing the band gap bowing from experimental alloy
data, the $GW$ estimate of the BAs direct band gap (5.5
eV),\cite{surh} which corrects LDA errors, is a more appropriate value
to use to calculate the bowing. Using the experimental alloy data from
Ref.~\onlinecite{geisz}, the $GW$-estimated band gap for BAs (5.5 eV),
and the experimental value of 1.42 eV for GaAs, Geisz
estimates\cite{geisznote} the revised bowing parameter to be $\sim$3.5
eV in excellent agreement with the theoretical prediction.

The bowing of B$_x$Ga$_{1-x}$As is much smaller than the bowing for
GaAs$_{1-x}$N$_x$ alloys in the same range. Accordingly, we predict
that addition of boron to GaAs will {\em raise} the gap. For example,
the predicted gaps for 2\%, 3\%, 5\% or 10\% concentrations of boron,
the band gaps would be 1.43, 1.44, 1.46, and 1.51 eV,
respectively. Thus, the addition of boron to GaAs will not lead to a 1
eV gap alloy as seen with nitrogen addition. We also find that the
bowing is composition independent. This implies that the effects of
boron on the band gap in GaAs are much less pronounced than those of
nitrogen. This is not completely unexpected, however, as it has been
previously observed that the bowing is smaller in mixed cation systems
than in mixed anion systems.\cite{shw-bowing}

We have also calculated the band gap of the BGaAs$_2$ CuPt-ordered
compound finding that it is 0.12 eV below the 50\%--50\% random alloy
modeled via the SQS16.

\subsection{Enthalpy of mixing and stability of superlattices}
\label{sec:deltaH}
Table \ref{deltaH} gives the formation enthalpies
[Eq.~(\ref{eq:deltaH})] for (001) {\em ordered superlattices} of BAs/GaAs and
the mixing enthalpies of B$_x$Ga$_{1-x}$As {\em random alloys}. The
formation energy of ordered superlattices of orientation $\hat G$ and
period $n$ is
\begin{equation}
\label{eq:superH1}
\Delta H (n,\hat G)=E(A_nB_n;\hat G)
-\frac{1}{2}(E_A+E_B)
\end{equation}
can be seperated into two contributions:\cite{dandrea} (i) The
consituent strain energy, i.e. the energy of the infinite period
superlattice. This is the strain energy associated with deforming pure
A and pure B into the in-plane lattice constant $\overline a$ of the
superlattice, and relaxing them in the direction $\hat G$. (ii) The
interfacial energy, i.e., the difference between $\Delta H(n,\hat G)$
and the consituent strain energy, is defined by
\begin{equation}
\label{eq:superH2}
\Delta H (n,\hat G)=
\frac{2I(n,\hat G)}{n}+\Delta E_{CS}(\overline
a,\hat G).
\end{equation}
We calculated $\Delta E_{CS}(\overline a,\hat G)$ for BAs and GaAs,
deformed to the average lattice constant $\overline a$ along $\hat
G=(001)$. This gave 148 meV/atom. From $\Delta H (n,\hat G=001)$ of
Table \ref{deltaH} and $\Delta E_{CS}$ we calculated the interfacial
energy $I$ for $n=1$, 2, and 4. We found $I=19$, 23, and 26 meV,
respectively. While a larger superlattice period $n$ would be required
to determine the converged value of $I$, we see that the BAs-GaAs
interface is {\em repulsive}. This is why $\Delta H (n,\hat G)$ of
Table \ref{deltaH} {\em decreases} with $n$: larger $n$ reduces the
{\em proportional} effect of the interfacial repulsion.

So far we dealt with (001) superlattices. The monolayer
(BAs)$_1$/(GaAs)$_1$ (111) is particularly interesting since in other
III--V alloys (e.g., GaInP$_2$) it appears as a spontaneously ordered
stucture during alloying.\cite{azreview} We find a very high $\Delta H
(CuPt)$ of 108 eV/atom, suggesting thermodynamic instability.

\begin{figure}[tbp] \centering
\includegraphics[width=\linewidth]{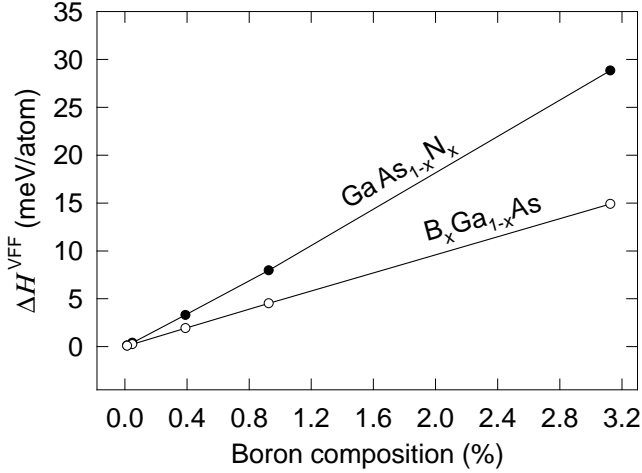}
\caption{Calculated VFF mixing enthalpies for dilute GaAs$_{1-x}$N$_x$
alloys and dilute B$_x$Ga$_{1-x}$As alloys. The mixing enthalpy is
substantially smaller for boron alloys.}
\label{mixingenthalpy}
\end{figure}

For the random alloys in Table \ref{deltaH}, all of the excess
enthalpies are positive. We compare the mixing enthaplies as obtained by
the LDA (which includes both size-mismatched-induced strain effects
and charge-tranfer ``chemical'' effects) and the valence force field
(VFF) method (which includes only strain effects). The trends and
magnitudes are similar, showing that the strain effects dominate
$\Delta H$.
In Fig.\ \ref{mixingenthalpy} we compare the VFF-calculated mixing
enthalpy of nitrogen in GaAs and contrast it with that of boron in
GaAs,
showing that the mixing enthalpy of GaAs$_{1-x}$N$_x$ alloys is much
higher than for B$_x$Ga$_{1-x}$As alloys. These results indicate that
it may be easier to alloy boron with GaAs. This implies that the
range of {\em bulk} B$_x$Ga$_{1-x}$As alloy compositions could be
substantially larger than for the nitride alloy GaAs$_{1-x}$N$_x$. In
{\em epitaxial} growth experiments, the alloy solubility can
dramatically exceed that in {\em bulk} experiments for reasons
explained in Ref.\ \onlinecite{sbz}.

\subsection{Quasi-localized electronic states}
\label{sec:loc}
\begin{figure}[tbp]
\includegraphics[width=\linewidth]{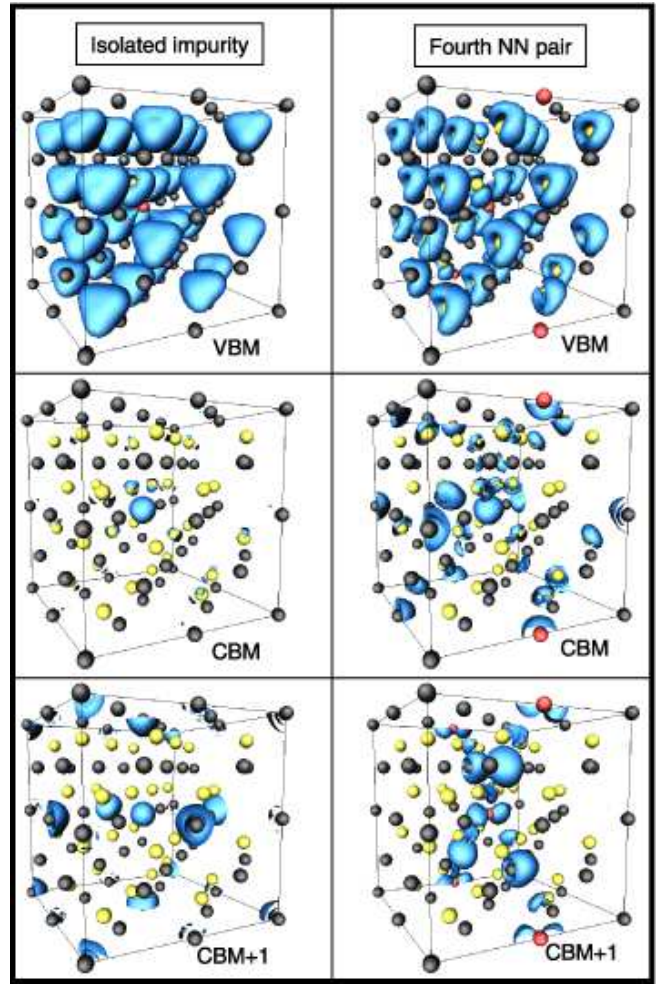}
\caption{(color) Isosurfaces of the wavefunction squared for a single
impurity and for a fourth nearest neighbor pair. Boron atoms are red,
gallium black, and arsenic yellow. The isosurface value (blue) is 0.01
e/\AA$^3$. In both cases, the VBM states are completely delocalized
and are primarily As-derived. The lower conduction band states are
strongly concentrated around the boron atoms.}
\label{localized}
\end{figure}

We find that the incorporation of boron into the GaAs host material
has little effect on the state at the VBM but that the lower
conduction band states are strongly peturbed. The square of the
wavefunctions for the VBM and the lowest two conduction band states
for isolated boron in GaAs are shown in the left hand side of Fig.\
\ref{localized}. The first column shows the states for an isolated
boron atom in the center of a supercell of GaAs. The VBM is completely
delocalized and looks very much like the VBM state of pure GaAs except
for a region near boron impurity where the wavefunction is extended
towards the boron atom. This distortion is precisely what one would
expect based on the strong coupling of boron $p$ and arsenic $p$
states at the VBM in BAs, as discussed in Secs.\ \ref{secbands} and
\ref{secrho}. In contrast to the VBM, the CBM state shows a
significant localization {\em around} the boron atom. The CBM shows
{\em long range} delocalization, but the majority of the wavefunction
is concentrated near the boron atom. For the second lowest conduction
band state (CBM+1), the situation is similar except the wavefunction
is concentrated around a small number of gallium atoms as well. The
dual character of the conduction band states (extended at long range
but concentrated around the boron atoms in the short range) indicates
that the states are {\em resonant} in the conduction band and are not
localized states {\em inside the gap}. The conduction band states
could be considered as boron-perturbed bulk GaAs conduction band
states.

The second column of Fig.\ \ref{localized} shows the same states (VBM,
CBM, CBM+1) for a fourth neighbor {\em pair} of boron atoms inside a
supercell of GaAs. A statistically random distribution of boron atoms
would result, among others, in pairs. In the case of GaAs$_{1-x}$N$_x$
alloys, the presence of nitrogen pairs can result in localized
impurity states inside the gap.\cite{xiao} Calculations were performed
for the five symmetrically-unique pair arrangements in the 64 atom
supercell as described in Table~\ref{pairs}.  Qualitative features of
wavefunction localization were the same for all of the pairs, and only
the most representative case, the fourth nearest neighbor pair, is
shown in the figure. The features of the pair states are similar to
those of the isolated impurity discussed above---that is, the VBM is
mainly an As-derived, delocalized state while the lower CB states are
concentrated around the boron atoms.  Again, there is some effect on
the wavefunction very near the boron atom. The CBM state is
concentrated around the boron atom in the middle of the cell and
around the boron and gallium atoms on the cell edges. The next lowest
conduction band state shows similar features and the wavefunction is
restricted primarily to the (200) plane of the cell.

The short-range ``localization'' effects of boron incorporation into
GaAs appear to be similar to those seen in GaAs$_{1-x}$N$_x$ alloys,
resulting mainly in ``dual character'' conduction band states that are
still extended at long range but are localized around the boron
atoms. However, it appears that the perturbation of boron on the
near-gap states of GaAs is gentler than that of nitrogen as none of
the pairs cause states inside the gap. This is another instance (as
with band gap bowing and band offsets) where boron is less
``intrusive'' than its first-row cousin, nitrogen.

\section{Conclusions}
This paper explored the effects of alloying a conventional III--V
compound, GaAs, with boron. Very little has been known, experimentally
or theoretically about boride III--V semiconductor alloys, but boride
semiconductor alloys are generating renewed interest now that epitaxial
techniques have made it somewhat easier to fabricate them. In order to
understand the fundamentals of this new class of possible alloys. We
have studied zinc-blende BAs and its alloy with GaAs.

For BAs we find: (i) The bonding in BAs is much more covalent than in
the rest of the III--As family or other III--Vs. This is due to the
similar electronegativities of boron and arsenic. (ii) Consequently,
BAs has a nearly symmetric, ``double-hump'' bond charge density,
similar to silicon. The small asymmetry that does exist in the bonding
charge density actually favors {\em boron}. That is, charge is drawn
toward the boron atom. In this sense, boron behave more like the {\em
anion} than the {\em cation} in BAs.  (iii) The band structure of BAs
is more reminiscent of silicon than most other III--V compounds. This
is a consequence of (1) the small repulsion between bonding and
antibonding $p$ states and (2) the large repulsion of the arsenic and
boron $s$ states. (iv) The LDA errors in the band gaps are
approximately constant in the III--As series BAs, AlAs, GaAs, and
InAs. (v) The band offsets of BAs/GaAs and BAs/AlAs are
unexpected. The VBM of BAs is 0.39 eV higher than AlAs and only 0.19
eV below GaAs. The main reason for this is the strong hybridization of
both the cation and anion states at the VBM.

For the B$_x$Ga$_{1-x}$As alloys, we find that: (i) The bond angles
are highly distorted near the boron sites. This accomodates the Ga--As
and B--As bonds to stay close to their ideal bulk values. As a result,
the bond length distribution in a random alloy exhibits the typical
bimodal behavior where each mode is near the individual ideal bulk
values. (ii) The bowing parameter of B$_x$Ga$_{1-x}$As for low
concentrations of boron is 3.5 eV, much smaller than for
GaAs$_{1-x}$N$_x$ alloys in the same composition range. Consequently,
unlike GaN, alloying GaAs with small amounts ($<10\%$) of BAs {\em
increases} the gap. (iii) The enthalpies of mixing indicate that the
bulk solubility of boron in GaAs may be significantly higher than
nitrogen in GaAs, possibly increasing the composition range over which
boride alloys may be formed. (iv) The wavefunction of the VBM is
completely delocalized and retains the character of pure GaAs. The
lower conduction band states exhibit a ``semi-localized''
behavior---i.e. the states are strongly concentrated around the boron
atoms, but far away they are extended states.

In summary, the behavior of B$_x$Ga$_{1-x}$As alloys is qualitatively
different from GaAs$_{1-x}$N$_x$ alloys. The perturbation of boron in
GaAs is much ``gentler'' on most features of the electronic structure
of GaAs than is nitrogen. If the same behavior is manifest for other
boride III--V alloys, boron offers a new class of III--V alloys to be
explored for novel behavior and possible device applications. Boron
may also play an important role in alloys where it could be used as a
relatively benign component which is added to lattice match the alloy
to a given substrate.

\section{Acknowledgements}
Supported by DOE-SC-BES-DMS under contract No.\ DE-AC36-99-GO10337. We
gratefully acknowledge S.-H. Wei for many helpful discussions
regarding the LAPW method and a critical reading of the manuscript,
and J. F. Geisz for useful discussions regarding the experimental data
and techniques as well as a critical reading of the manuscript.


\end{document}